\documentclass[acmsmall]{acmart}
\AtBeginDocument{%
  }

\acmJournal{JACM}
\acmVolume{0}
\acmNumber{0}
\acmArticle{0}
\acmMonth{0}

\usepackage[most]{tcolorbox}
\usepackage{tcolorbox}
\usepackage{todonotes}

\usepackage{fontawesome5}

\usepackage{pdflscape}
\usepackage{rotating}    
\usepackage{booktabs}    
\usepackage{makecell}    
\usepackage{adjustbox}   
\begin{document}

\title{Investigating VR Accessibility Reviews for Users with Disabilities: A Qualitative Analysis}

\author{Yi Wang}
\affiliation{%
 \institution{School of Information Technology, Deakin University}
 \city{Geelong}
 \state{VIC}
 \country{Australia}}
 \email{xve@deakin.edu.au}

 \author{Chetan Arora}
\affiliation{%
  \institution{Faculty of Information Technology, Monash University}
 \city{Melbourne}
 \state{VIC}
 \country{Australia}}
  \email{chetan.arora@monash.edu}

\author{Xiao Liu}
\affiliation{%
  \institution{School of Information Technology, Deakin University}
 \city{Geelong}
 \state{VIC}
 \country{Australia}}
 \email{xiao.liu@deakin.edu.au}

\author{Thuong Hoang}
\affiliation{%
  \institution{School of Information Technology, Deakin University}
 \city{Geelong}
 \state{VIC}
 \country{Australia}}
\email{thuong.hoang@deakin.edu.au}

\author{Zhengxin Zhang}
\affiliation{%
   \institution{School of Design, Inner Mongolia Normal University}
  \city{Hohhot}
  \country{China}}
\email{robinzhang2015@iCloud.com}

  \author{Henry Been-Lirn Duh}
\affiliation{%
  \institution{School of Design, The Hong Kong Polytechnic University}
 \city{Hong Kong}
 \country{China}}
  \email{henry.duh@polyu.edu.hk}

  \author{John Grundy}
\affiliation{%
  \institution{Faculty of Information Technology, Monash University}
 \city{Melbourne}
 \state{VIC}
 \country{Australia}}
  \email{john.grundy@monash.edu}

\renewcommand{\shortauthors}{Wang et al.}

\begin{abstract}
Accessibility reviews provide valuable insights into both the limitations and benefits experienced by users with disabilities when using virtual reality (VR) applications. However, a comprehensive investigation into VR accessibility for users with disabilities is still lacking. To fill this gap, this study analyzes user reviews from the Meta and Steam stores of VR apps, focusing on the reported issues affecting users with disabilities. We applied selection criteria to 1,367,419 reviews from the top 40, the 20 most popular, and the 40 lowest-rated VR applications on both platforms. In total, 1,076 (0.078\%) VR accessibility reviews referenced various disabilities across 100 VR applications. These applications were categorized into Action, Sports, Social, Puzzle, Horror, and Simulation, with Action receiving the highest number of accessibility related-reviews. We identified 16 different types of disabilities across six categories. Furthermore, we examined the causes of accessibility issues as reported by users with disabilities. Overall, VR accessibility reviews were predominantly under-supported.
\end{abstract}




\maketitle

\section{Introduction}

Accessibility is a critical requirement for users with disabilities across multiple platforms, including mobile applications~\cite{Alomar21,fazzini2022characterizing}, websites~\cite{Miranda22, Inal22}, and virtual reality (VR)~\cite{South24, Naikar24}. There is an increasing focus on methods, tools, and frameworks designed to help software practitioners create inclusive software applications throughout the software development life cycle~\cite{Haggag22, Manca23, Shinohara20, Regimbal24}. Despite these advancements, existing literature still provides insufficient guidance on integrating accessibility requirements and needs into the software development life cycle of immersive technologies, such as VR~\cite{dudley2023inclusive}. Furthermore, the World Wide Web Consortium (W3C) has released a draft of the XR Accessibility User Requirements (XAUR) to guide virtual reality (VR) practitioners in making immersive applications accessible to a diverse range of users with disabilities. However, many VR software developers, particularly those from the gaming and visual effects industries, often lack expertise in the requirements engineering (RE) practices common in the software development life cycle~\cite{aleem2016game}. Additionally, traditional requirements engineering methods do not sufficiently address more complexity inherent in VR projects, which involve entities such as 3D models, virtual environments, lighting, and audio components~\cite{Karre24, WANG2025107609}. Previous studies indicated that the RE practices in VR projects differ substantially from those in traditional software projects (e.g., web and mobile applications)~\cite{Wang25Understanding}. For example, common requirements elicitation methods, such as personas~\cite{zhang2023personagen}, are typically not employed in VR apps~\cite{WANG2025107609}. Thus, traditional approaches to identifying accessibility requirements may not fully align with the VR software development life cycle.


%
VR accessibility is a crucial factor influencing the acceptance and inclusivity of users with disabilities~\cite{pladere2022inclusivity}. In recent years, researchers have increasingly focused on VR accessibility issues. For instance, one study examines the challenges that users with visual impairments or low vision face when interacting with objects in virtual environments~\cite{Zhao19}, while another explores methods for visualizing audio features, such as volume, duration, and spatial location, to support users with hearing impairments~\cite{Li22}. The XAUR guidelines categorize disabilities into six types: auditory, cognitive, neurological, physical, speech, and visual. They provide a preliminary overview of typical disabilities, alongside their corresponding accessibility requirements. However, there are few comprehensive investigations on VR accessibility in practical VR software projects~\cite{Naikar24}. Studies on accessibility reviews mainly focus on mobile applications, such as those available in the IOS and Android stores~\cite{Oliveira23, Alomar21, Manoel21}. Incorporating accessibility reviews can enhance software practitioners' understanding of the needs of users with disabilities. However, many software practitioners may not prioritize accessibility in their software projects~\cite{Bi22}. Unlike mobile apps, there is a notable lack of studies evaluating the current state of VR accessibility reviews across various VR app stores~\cite{Wang25Understanding}.

To address this gap, we present an investigation in which we identified and analyzed user reviews on accessibility, reporting both the feedback and the causes of accessibility issues. We collected a total of 1,367,419 user reviews from the top 40, the most popular 20, and the lowest-rated 40 on the Meta and Steam stores. Our three-step data collection process includes user review collection, filtering through string matching, and manual inspection. Finally, we extracted a total of 1,076 accessibility reviews. The main goals of our study are to: (a) investigate the number of accessibility reviews mentioned by disabled users; (b) investigate the different types of disabilities mentioned by users with disabilities; (c) investigate the various types of disabilities across different types of VR applications; and (d)analyze the causes of accessibility issues for users with different types of disabilities. The main contributions of this paper are:

\begin{itemize}
    \item We collect more than 1.3 million reviews from the top 40, the most popular 20 and the lowest-rated 40 applications on the Meta and Steam stores. From these, we extract 1,076 reviews related to accessibility. To the best of our knowledge, this constitutes the first comprehensive investigation of VR accessibility reviews to date. 
    \item We present firsthand evidence from accessibility reviews contributed by users with disabilities. In total, we categorized feedback as positive, negative, and neutral, and identified 16 types of disabilities. 
    \item We present the primary reasons for accessibility issues as reported in user reviews. We found that users with motion sickness and hearing impairments most frequently mentioned these issues, with the largest group being those affected by motion sickness. Conversely, other disabilities were less commonly represented and identified fewer accessibility-related concerns.
    \item We categorize user attitudes towards VR accessibility feedback as positive, negative, and neutral. 
\end{itemize}

\textbf{Paper Organization. }Section \ref{rw} summarizes the related work. Section \ref{sd} introduces our research questions, the dataset identification process with VR accessibility reviews, and the design of the research method. Section \ref{res} presents the key findings. Section \ref{dis} discusses the main findings and future research recommendations. Section \ref{limi} outlines the threats to validity, and Section \ref{con} concludes this work.


\section{Related Work}\label{rw}

Ensuring that a wide range of users, including those with varying abilities, can access different applications is essential \cite{heron2013open}. Previous studies primarily focused on accessibility issues in user reviews of mobile applications \cite{Oliveira23, Manoel21, Eler19, Aljedaani22, Yan19}. Research on VR accessibility primarily focused on prosocial guidelines \cite{heilemann2021accessibility, Abeele21}, accessibility tools and technologies \cite{Zhao19, Li22, Li23, MP18}, and the accessibility evaluations of existing VR applications \cite{Naikar24}. To the best of our knowledge, no prior study has investigated VR accessibility reviews across multiple VR applications.


\subsection{Accessible Virtual Reality}

Unlike mobile applications, VR applications place a strong emphasis on delivering immersive user experiences. Achieving high-quality immersion depends on several resources in virtual environments, such as audio assets \cite{Li22}, models \cite{Reinschluessel19}, user interfaces \cite{Kim17}, interactions \cite{MS19}, and comfort \cite{Bajorunaite22}. Nevertheless, users with disabilities and specific accessibility challenges may require appropriate support to ensure that they can use VR applications. Additionally, the user base includes special user groups who are not disabled users, such as the elderly and children \cite{Wu24, Jin24}. They also require VR applications that are easy to access and understand.

In recent years, many studies have focused on designing VR systems to assist users with accessibility challenges or users with disabilities. Li et al. \cite{Li24} developed a multisensory VR game system prototype based on cognitive psychology paradigms, designed to assist elderly users with limited mobility, particularly those who cannot stand, in engaging with VR games. Wu et al. \cite{Wu24} indicated that factors contributing to interaction difficulties for elderly users included the requirement for simultaneous interaction with both hands, age-related variability, and differences in rotating viewpoints and target heights. Additionally, many studies focused on how to use VR to alleviate the symptoms of autism in children. For example, Delano and Snell \cite{delano2006effects} noted that social stories provide clear visual narratives to explain social situations and concepts, which can help children with autism. Zhang et al. \cite{zhang23} considered that integrating social stories into VR can provide more interactive, immersive experiences and flexible interventions. Boyd et al. \cite{body22} proposed a web-based VR system that can support the coping strategies of multiple stakeholders, including autistic adults, children with autism, and the legal guardians of children with autism. This VR system is designed with various virtual environment options, including distraction isolation, movement, remaining still, and engagement. Su and Yan \cite{Su23} designed a VR game aimed at helping children and stakeholders collaborate to prevent dark phobia. Furthermore, many studies focused on developing tools to support users with disabilities through VR applications. Zhao et al \cite{Zhao19} developed a toolkit that enhances the visual and auditory experience to improve the experience of users with low vision in virtual environments. Similarly, Li \cite{Li22, Li23} et al. proposed a tool (\textit{SoundVizVR}) that uses auditory feature indicators to visualize the loudness, duration, and location of sound sources in virtual environments and employs indicators to provide additional information about sounds. They also applied the \textit{SoundVizVR} plugin in VR game development practices. However, numerous challenges have emerged in development practices. For instance, VR practitioners may struggle to find suitable accessibility plugins in the Unity store. Independent game development teams often lack professional accessibility experts, and similar challenges persist in traditional software development practices \cite{Bi22}. Moreover, a previous study found that free VR applications on the Meta store still do not provide key accessibility features, such as color blindness, text to speech, and speech to text capabilities \cite{Naikar24}.

\subsection{Accessibility in User Reviews}

Accessibility reviews are an important approach to understanding the needs of users with disabilities \cite{Bi22, Wang25Understanding}. Currently, accessibility reviews are primarily conducted for mobile applications on the Google Play Store or Apple Store. However, even fully developed applications still have accessibility issues \cite{Eler19}. As an example, only 1.24\% of users reported accessibility issues to mobile app stores. A previous study suggested that the terms used in accessibility reviews may not always align with accessibility guidelines. Thus, categorizing user terminology to match accessibility guidelines can help designers and developers quickly infer the required accessibility features. Santos et al. \cite{DS24} investigated accessibility reviews after software updates and reported that updates might lead to incompatibilities with accessibility features. Aljedaani et al. \cite{Aljedaani23} found that most deaf or hard-of-hearing students encounter barriers when using Learning Management Systems (LMS). 31\% of these systems did not comply with accessibility principles. Park et al. \cite{Park14} found that the main difficulties for users with visual impairments are the speed of text input and the design of the application, which often lacks accessibility. Furthermore, many studies focused on the automatic identification of accessibility reviews \cite{Swearngin24, Alomar21, Alshayban22, Chen22}. Alshayban and Malek \cite{Alshayban22} proposed an automatic detection technique for accessibility issues (\textit{AccessiText}), aimed at identifying problems caused by text scaling assistive services in Android and iOS applications. AIOmar et al. \cite{Alomar21} used 5,326 accessibility reviews to evaluate their designed model. Swearngin et al. \cite{Swearngin24} proposed a system that automatically generates accessibility reports, which 19 accessibility engineers found satisfactory regarding the auto-generated accessibility reports and prioritized accessibility issue lists. Additionally, Oliveira et al. \cite{Oliveira23} employed topic modeling techniques to automatically generate several topics based on accessibility reviews from users with visual impairments. An empirical investigation found that 88.99\% of applications have accessibility issues \cite{Chen22}. However, to the best of our knowledge, no studies have investigated VR accessibility reviews specifically associated with users who have disabilities.


\section{Study Design}\label{sd}

In this study, we analyzed VR accessibility reviews from users with disabilities on the Meta\footnote{https://www.meta.com/en-gb/experiences/} and Steam store\footnote{https://store.steampowered.com/vr/}. We then categorized these VR accessibility reviews as positive, negative, and neutral based on different types of disabilities and VR applications.

\subsection{Research Questions}

We framed our study on the following four key research questions (RQs):

\textbf{RQ1:} \textit{How many accessibility reviews are mentioned by users with disabilities?} To answer this research question, we categorized six different types of VR applications into the following categories: Action, Sport, Social, Puzzle, Horror, and Simulation. We then analyzed the distribution of positive, negative, and neutral accessibility reviews across these categories and summarized the sample sizes for each.



\textbf{RQ2:} \textit{Which disabilities are discussed in accessibility reviews?} For this research question, we summarized the number of different types of disabilities and specifically highlighted the counts of positive, negative, and neutral reviews. 


\textbf{RQ3:} \textit{What is the categorization of accessibility reviews of VR applications?} Our goal is to examine the accessibility reviews referencing disabilities across different types of VR applications. These results provide insights into the distribution of accessibility reviews among categories and help identify which types of VR applications receive more accessibility reviews.

\textbf{RQ4:} \textit{What are the reasons underlying accessibility issues for users with disabilities?} For this research question, we analyzed all accessibility reviews and identified the causes of accessibility issues. These results are useful for VR practitioners seeking to understand the accessibility challenges reported by users with disabilities.








\begin{figure*}[t]
    \centering
    \includegraphics[width=\linewidth]{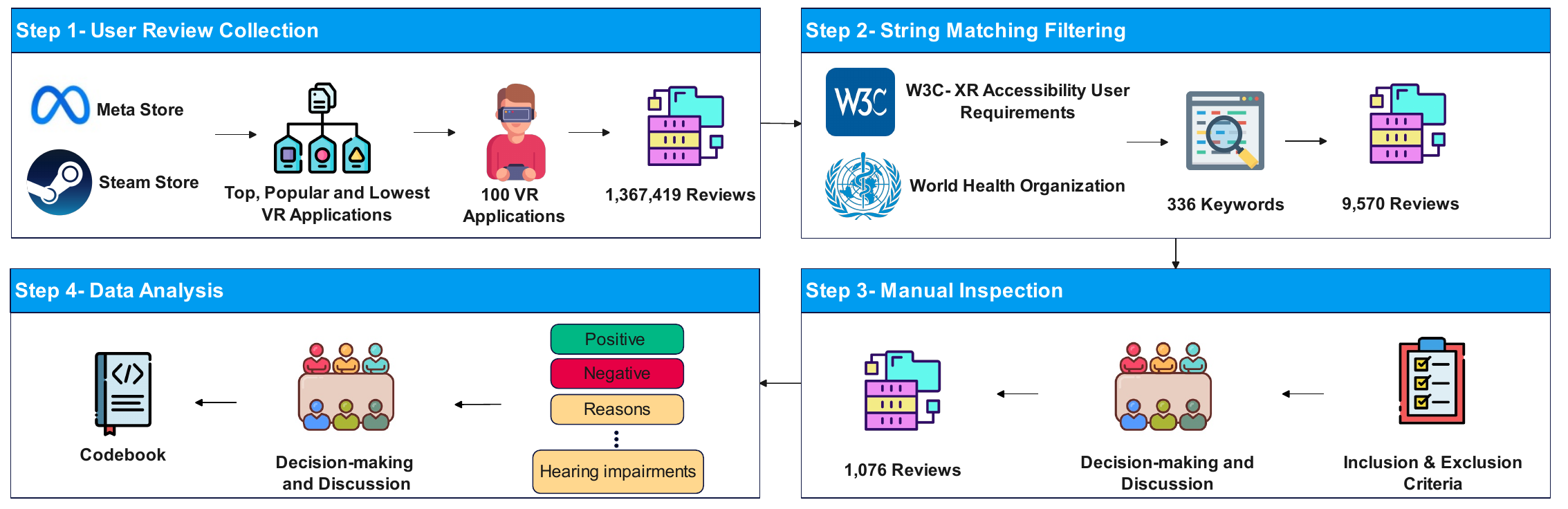}
    \caption{Overview of each step of our sampling process.} 
    \label{fig:enter-fig1}
\end{figure*}

\subsection{Sampling Method}

We employed a purposive sampling method to select reviews focusing on VR accessibility from the top 20 (highest rated) VR applications on the Meta and Steam stores \cite{tongco2007purposive}. To achieve a more diverse sample, we also included the most popular 10 VR applications on both platforms. In addition, We selected reviews from the 20 lowest-rated VR applications on the Meta and Steam stores \cite{ullah2023exploring}. The most popular VR applications were identified using the ``popular'' filters on the Meta store and by ranking VR applications based on player counts on the Steam store. The top and the lowest-rated VR applications were selected using user rating filters on both platforms. To avoid overlap between the top 40 and the most popular 20 and the most lowest-rated 40 VR applications, we reviewed titles appearing in both lists for each platform and found no overlaps. In contrast, overlaps were identified between the Meta and Steam Stores, including one application appearing among the top-ranked titles and four among the most popular titles. No overlaps were found among the lowest-rated applications. Given that Meta and Steam cater to different VR headsets and user preferences, all overlapping applications from both platforms were retained. In total, we selected 50 VR applications from each platform, resulting in a dataset of 100 applications.





The sampling process was conducted in three steps. Figure \ref{fig:enter-fig1} shows the four main steps of our procedure. The details of each step are presented as follows:

\textit{(1). User Reviews Collection.} We collected 1,367,419 reviews from the 100 VR applications selected on the Meta and Steam Stores. On the Meta Store, the top 20 VR applications accounted for 57,530 reviews, the 10 most popular applications for 73,294 reviews, and the lowest-rated 40 VR applications for 3,530 reviews. On the Steam Store, the top 20 VR applications accounted for 774,214 reviews, the 10 most popular applications for 458,549 reviews, and the lowest-rated 20 applications for 302 reviews. For both the top and most popular categories on Steam, Action VR applications received 444,616 reviews, Horror VR applications 371,033 reviews, Social VR applications 242,510 reviews, Puzzle VR applications 15,294 reviews, Sports VR applications 105,788 reviews, and Simulation VR applications 53,497 reviews. Only the contents of the reviews were captured; no details of the reviewer were included. All VR applications on the Meta and Steam stores were confirmed by the end of February 2025, and all data were collected by the end of April 2025.


\textit{(2). String-Matching Filtering.} We employed string-matching filtering based on specific keywords to identify reviews mentioning users with accessibility challenges. To ensure that no relevant data was missed, we employed fuzzy matching during the keyword search process. Given that the term `users with accessibility challenges' covers a broad range of user groups, we referred to the XR Accessibility User Requirements (XAUR)\footnote{https://www.w3.org/TR/xaur/}. XAUR identifies various types of accessibility needs and requirements as keywords. Based on these, we designed 210 specific keywords, including accessibility needs and various types of disabilities, such as `immersive personalization', `user preferences', `change speed', `time limits', `motion sickness', and so on. Table \ref{tab:vr-acc-keywords-by-theme} and \ref{tab:vr-acc-keywords-by-theme2} in the appendix show all the keywords. To further expand the set of disability-related keywords, we referred to the World Health Organization (WHO)\footnote{https://www.who.int/standards/classifications/classification-of-diseases} at ICD-11 for Mortality and Morbidity Statistics\footnote{https://icd.who.int/browse/2024-01/mms/en}, and International Classification of Functioning, Disability and Health\footnote{https://icd.who.int/browse/2024-01/icf/en}. From these sources, we derived an additional 84 keywords, including four abbreviations. All keywords were aligned with the six categories of disabilities defined by XAUR, including \textit{`Auditory Disabilities'}, \textit{`Cognitive Disabilities'}, \textit{`Neurological Disabilities'}, \textit{`Physical Disabilities'}, \textit{`Speech Disabilities'}, and \textit{`Visual Disabilities'}. Table \ref{tab:who-disability-keywords} in the appendix shows all the keywords. After a final review of both the XAUR and WHO glossaries, we confirmed that we had reached saturation, with no further new keywords emerging. In total, 294 specific keywords were defined. Using these keywords for string-matching, we extracted 9,570 accessibility reviews.






\textit{(3). Manual Inspection.} String-matching may result in false positive reviews, reviews that do not meet the criteria, and reviews containing malicious content. Four experienced researchers manually reviewed the datasets to ensure they met the accessibility requirements for users with disabilities. One experienced researcher examined the datasets from the 40 lowest-rated VR applications on the Meta and Steam Stores. Inclusion and exclusion criteria were defined to standardize the decision-making process.

Our inclusion criterion defined that the final sample include accessibility reviews from users with disabilities, such as \textit{ ``I am hearing impaired so I cannot hear what people say in the game which makes them not pay any attention to me or just flat out quit.''} Meanwhile, the reviews in which users self-identified as having a disability, such as \textit{``I was sick for almost 2 weeks as my eyes got sensitive to light, my ears to sound, I had headaches, fatigue and lost my balance while washing the dishes [in VR].''} The reviews also included feedback on the assistive technologies used by the user with a disability or on their experience with the VR application, such as \textit{``...VRC-CC [App Name - Creator Companion] which adds captions for those hard of hearing to still actually be able to play this game I certainly won't be playing this game much longer.''} Additionally, considering that users with disabilities might be reluctant to speak out and may instead discuss accessibility issues from a third-person perspective \cite{hydock2020unhappy}, three researchers jointly reviewed and discussed the quality and reliability of the comments. Comments were included if they provided sufficient information.

Our exclusion criterion defined that reviews must be excluded in eight cases: 1. It was incomplete and lacked an accompanying explanation, such as \textit{``Terrible for people who get motion sickness.''}. 2. Reviews on behavioral preferences provided by users without disabilities, such as \textit{``I am left-handed. It is not a curse, disease or disability. I just am.''} 3. Reviews that provided only limited accessibility suggestions for users with disabilities, where the reviewers themselves were not disabled, such as \textit{``you never regret it (Motion sickness free).''} 4. Reviews that did not clearly indicate whether assistive technology could help users, such as \textit{``Also much appreciation to all of the accessibility settings that Valve added.''} 5. If users did not specify the type of disability they had, such as \textit{``I'm 60 years old with disabilities.''}. 6. Some terms describing disabilities could also be used to describe 3D avatars or game objectives in VR, such as \textit{``...[Avatar] I've broken a bone.''} 7. Some reviews from users without disabilities mentioned accessibility requirements in VR, such as \textit{``I can only stand and play the game, can't sit down.''}  8. Duplicate comments and malicious remarks toward users with disabilities were excluded from the database.

During the manual inspection, if researchers were uncertain about including a particular review, the decision was made through a joint discussion. A more experienced researcher then reviewed the final datasets and made the final inclusion decisions in collaboration with the other four researchers. In total, 1,076 reviews were collected, of which 46 were from the 40 lowest-rated VR applications on the Meta and Steam Stores.






\subsection{Thematic Analysis }

We employed manual analysis for three primary reasons: (1) the volume of accessibility reviews collected was moderate; (2) VR applications on the Steam Store offer only “recommend” or “not recommend” options without specific star ratings; and (3) although the Meta Store provides 1–5 star ratings, it lacks an open API, preventing the collection of relevant rating data. Thus, we employed a thematic analysis approach combining both inductive and deductive methods \cite{nowell2017thematic, braun2006using} to explore accessibility issues in VR applications for users with disabilities. Thematic analysis facilitates the identification of patterns in text while considering contextual factors, such as positive, negative, and neutral sentiments, as well as their underlying causes \cite{kiger2020thematic}. We analyzed the accessibility reviews using MAXQDA\footnote{https://www.maxqda.com/}. After sampling, the reviews were classified as positive, negative, and neutral. Positive accessibility reviews indicated overall satisfaction with the VR application and the absence of accessibility barriers. Negative reviews indicated overall dissatisfaction and the presence of accessibility barriers. Neutral reviews indicated overall satisfaction but also highlighted one or more accessibility issues.

To analyze the coded data, we first adopted a deductive approach, identifying accessibility reviews according to the issues outlined in the XAUR guidelines. These guidelines define six main categories of disabilities: auditory, cognitive, neurological, physical, speech, and visual. In addition, the XAUR guidelines specify 19 categories of accessibility-related user needs and requirements. For example, these include adjusting color for users with visual impairments and providing immersive semantics for users with hearing impairments, such as assistive navigation, location identification, object recognition, and interaction support. Based on the XAUR, we defined six categories of disability themes, each containing three sub-themes: positive, negative, neutral and reason. The researchers determined whether the coded content aligned with the themes defined in the XAUR, following the World Health Organization’s guidance on the definition of disability. For any disabilities not explicitly mentioned in the XAUR, we employed an inductive analysis to identify new themes. This led us to introduce an additional theme, `Other disabilities,' which also included positive, negative, and neutral sub-themes.

Furthermore, we categorized different types of VR applications into five themes, including `Sport', `Social', `Action', `Puzzle', `Horror' and `Simulation'. Each VR application may contain multiple category tags; for instance, an action VR applications might also include other tags such as FPG (first-person game), multiplayer, horror and so on. Therefore, we defined the most relevant category tags based on the VR application's official description, videos, and definitions from Wikipedia\footnote{https://www.wikipedia.org/}. These application themes included six major categories of disabilities, as well as `Other Disability' themes.



We coded the causes of accessibility issues based on the categorized VR applications, using `Negative' and `Neutral' sub-themes. This process aimed to identify which accessibility issues were mentioned by users with disabilities and to determine the factors that might affect their use of VR applications. Meanwhile, some reviews included developer responses, which also reflected the developers’ attitudes toward addressing accessibility issues. To ensure consistency in the coding process, the code lists underwent four iterative rounds of discussion and merging, with the final list confirmed during the fourth round. Coding, thematic development, and mapping were conducted by the first author in ongoing consultation with four researchers until consensus was reached.





   
  %
    
    


\section{Results}\label{res}

In this section, we present the results of our investigation. The results are organized by research question, with results reported for each corresponding analysis.

\subsection{RQ1. How many accessibility reviews are mentioned by users with disabilities?}

We identified 1,076 accessibility-related reviews, accounting for only 0.075\% (\textit{N = 1,367,419}) of all VR application reviews in the dataset. Accessibility reviews for the top 40, and the most popular 20, and the lowest-rated 40 VR applications on the Meta store (513 reviews) were fewer than those on Steam (563 reviews). Of the identified accessibility reviews, 408 (37.9\%) were positive, 484 (44.9\%) were negative accessibility reviews, and 184 (17.1\%) neutral accessibility reviews. VR applications classified in the action category received the highest number of accessibility reviews on both the Meta and Steam stores. Meanwhile, action VR applications on the Meta store received the highest number of positive feedback. Simulation VR applications received the fewest accessibility reviews on the Meta store, while Horror VR applications had none. On the Stream, Sport and Horror VR applications received the fewest accessibility reviews. In contrast, Action VR applications received the highest number of accessibility reviews on the Steam store, where negative feedback exceeded positive feedback. Tables \ref{tab:meta} and \ref{tab:steam} summarize the number of accessibility reviews for the top 40, the most popular 20, and the most lowest-rated 40 VR applications from the Meta and Steam stores. Across most VR applications, negative accessibility reviews were predominant. All categories of VR applications contained accessibility reviews associated with various types of disabilities.


For the 40 VR applications with the lowest ratings, accessibility was rarely mentioned in their reviews. On Steam, only 2 (0.66\%) negative accessibility reviews were reported by users with disabilities, compared to 44 (1.25\%) accessibility reviews on the Meta store. Of the Meta store reviews, 10 were positive accessibility reviews, 31 negative, and 5 were neutral. Overall, accessibility was mentioned infrequently in reviews of the lowest-rated VR applications.

\begin{table*}[t]
\centering
\caption{Meta Store: Accessibility requirements reviews of the top 20, the most popular 10, and the lowest-rated 20 VR applications.}
\label{tab:meta}
\begin{tabular}{lcccccccccc}
\toprule
\textbf{Category} & \multicolumn{2}{c}{\textbf{SAMPLE}} &\multicolumn{2}{c}{\textbf{POSITIVE}} &\multicolumn{2}{c}{\textbf{NEUTRAL}} & \multicolumn{2}{c}{\textbf{NEGATIVE}} \\
\cmidrule(r){2-3} \cmidrule(r){4-5} \cmidrule(r){6-7}\cmidrule(r){8-9}
& Reviews & Percent & Reviews & Percent & Reviews & Percent & Reviews & Percent \\
\midrule

Action & 295 &  57.5\% & 214 &72.5\% & 4 & 1.3\% & 77 & 26.1\% \\

Sport  & 61 & 11.8\% & 31 & 50.8\% & 10 & 16.3\% & 20 & 32.7\% \\

Social & 71 & 13.8\% & 8 & 11.2\% & 3 & 4.2\% & 60 & 84.5\% \\

Puzzle & 73 & 14.2\% & 37 & 50.6\% & 18 & 24.6\% & 18 & 24.6\%  \\

Horror & -- &--  & --  & -- & -- & --  & --  &  --   \\

Simulation & 13 & 2.5\% & 8 & 79.1\% & 1  & 4.16\%  & 4 & 16.6\% \\

\bottomrule
\end{tabular}
\end{table*}

\begin{table*}[t]
\centering
\caption{Steam Store: Accessibility requirements reviews of the top 20, most popular 10, and the lowest-rated 20 VR applications.}
\label{tab:steam}
\begin{tabular}{lcccccccccc}
\toprule
\textbf{Category} & \multicolumn{2}{c}{\textbf{SAMPLE}} &\multicolumn{2}{c}{\textbf{POSITIVE}} &\multicolumn{2}{c}{\textbf{NEUTRAL}} & \multicolumn{2}{c}{\textbf{NEGATIVE}} \\
\cmidrule(r){2-3} \cmidrule(r){4-5} \cmidrule(r){6-7}\cmidrule(r){8-9}
& Reviews & Percent & Reviews & Percent &  Reviews & Percent & Reviews & Percent \\
\midrule

Action & 232 &  41.2\% & 55 & 23.7\% & 80 & 34.4\% & 97 & 41.8\% \\

Sport  & 18 & 3.1\% & 4 & 22.2\% & 4 & 22.2\% & 10 & 55.5\% \\

Social & 207 & 36.7\% & 9 & 4.3\% & 37 & 17.8\% & 161 & 77.7\% \\

Puzzle & 23 & 4.0\% & 12 & 52.1\% & 6 & 26.0\% & 5 & 21.7\%  \\

Horror & 18 & 3.1\% & 1 & 5.5\% & 5 & 27.7\% & 12 & 66.6\%   \\

Simulation & 65 & 11.5\% & 29 & 44.6\% & 16 & 24.6\% & 20 & 30.7.0\% \\

\bottomrule
\end{tabular}
\end{table*}

\begin{tcolorbox}[
  colback=white,        
  colframe=black,       
  arc=3mm,              
  boxrule=1pt,        
  left=5pt, right=5pt,  
  top=5pt, bottom=5pt   
]

\noindent
\textbf{RQ1 Findings:} 

\begin{itemize}
    \item There was a very low number of accessibility reviews available overall for analysis, and among these, negative accessibility reviews predominated.
    
    \item Reviews varied by platform and VR application domain.
\end{itemize}

\end{tcolorbox}

\subsection{RQ2. Which disabilities are discussed in accessibility reviews?}

We identified 17 different types of disabilities mentioned in the VR accessibility reviews. Table \ref{tab:3} shows the number of VR accessibility reviews for all VR applications referencing disabilities. Among the identified issues, motion sickness is the most frequently reported accessibility issue. A total of 298 (35.9\%) reviews contained negative feedback, while 148 (17.8\%) reviews describe the VR applications enjoyable yet still associated with motion sickness. Among these, 89 reviews noted that prolonged use of the VR application helped users overcome motion sickness (see Table \ref{tab:3}). In contrast, 382 (46.1\%) reviews provided positive feedback, indicating that users with disabilities did not experience motion sickness with the current VR application but had encountered it with other VR applications.

For hearing impairments, accessibility reviews were predominantly negative (127; 91.3\%), with only two reviews (1.4\%) providing positive feedback and 10 (7.1\%) providing neutral feedback. For speech impairment, most reviews were negative (23; 58.9\%) or neutral (14; 35.9\%), with only two reviews (5.2\%) providing positive feedback. For epilepsy, reviews were more often negative (20; 60.6\%) than positive (8; 24.3\%), while five reviews (15.1\%) were neutral. For upper limb disabilities, reviews were evenly split between negative (6; 50\%) and positive (4; 33.3\%), with two reviews (16.6\%) classified as neutral.

Disabilities with fewer than ten accessibility reviews included autism spectrum disorder, blurred vision, lower limb disability, color blindness, social anxiety, attention deficit hyperactivity disorder, asthma, cerebral palsy, learning disability, and myopia. Some disabilities, such as asthma, cerebral palsy, learning disability, and myopia, were each mentioned only once, with only one positive review associated with cerebral palsy. Based on the XAUR’s six categories of disabilities and the WHO’s definitions, auditory disabilities included hearing impairments; cognitive disabilities included attention deficit hyperactivity disorder, autism spectrum disorder, and learning disability; neurological disabilities included motion sickness and epilepsy; physical disabilities included upper disability, lower limb disability, cerebral palsy, and asthma; and visual disabilities included blurred vision, myopia and color blindness. There is only one category for speech disabilities, which was classified as speech impairment.



\begin{tcolorbox}[
  colback=white,        
  colframe=black,       
  arc=3mm,              
  boxrule=1pt,        
  left=5pt, right=5pt,  
  top=5pt, bottom=5pt   
]

\noindent
\textbf{RQ2 Findings:} 

\begin{itemize}
    \item The majority of accessibility reviews from users with disabilities were predominantly negative, whereas only motion sickness reviews were most positive.
    \item There was difference in the number of accessibility issues reported by users with different types of disabilities, with hearing impairments and motion sickness accounting for the majority. 
\end{itemize}

\end{tcolorbox}


\clearpage
\begin{landscape}
\thispagestyle{empty} 

\begin{center}
\captionof{table}{Different types of disabilities of the selected VR applications.}
\label{tab:3}

\resizebox{\dimexpr\linewidth-3\baselineskip}{!}{%
  \begin{tabular}{lcccccccc} 
  \toprule
  \textbf{Disabilities}
    & \multicolumn{2}{c}{\textbf{SAMPLE}}
    & \multicolumn{2}{c}{\textbf{POSITIVE}}
    & \multicolumn{2}{c}{\textbf{NEUTRAL}}
    & \multicolumn{2}{c}{\textbf{NEGATIVE}} \\
  \cmidrule(r){2-3} \cmidrule(r){4-5} \cmidrule(r){6-7} \cmidrule(r){8-9}
  & Reviews & Percent & Reviews & Percent & Reviews & Percent & Reviews & Percent \\
  \midrule

  \multicolumn{9}{l}{\textbf{Auditory disabilities}} \\
  Hearing impairment                 & 136 & 11.5\% & 2   & 1.6\%  & 11  & 8.0\%  & 123 & 90.4\% \\

  \multicolumn{9}{l}{\textbf{Cognitive Disabilities}} \\
  \makecell[l]{Attention deficit\\hyperactivity disorder}
                              & 2   & 0.1\%  & 1   & 50.0\% & -- & --     & 1   & 50.0\% \\
  Learning disability         & 1   & 0.09\% & --  & --     & -- & --     & 1   & 100.0\% \\

  \multicolumn{9}{l}{\textbf{Neurological Disabilities}} \\
  Autism spectrum disorder    & 9   & 0.8\%  & 3   & 33.3\% & -- & --     & 6   & 66.6\% \\
  Cerebral palsy              & 1   & 0.09\% & 1   & 100.0\%& -- & --     & --  & --     \\
  Motion sickness             & 818 & 76.3\% & 382 & 46.1\% & 148& 17.8\% & 298 & 35.9\% \\
  Epilepsy                    & 33  & 3.0\%  & 8   & 24.3\% & 5  & 15.12\%& 20  & 60.6\% \\

  \multicolumn{9}{l}{\textbf{Speech impairment}} \\
  Speech impairment           & 39  & 3.5\%  & 2   & 5.2\%  & 14 & 35.9\% & 23  & 58.9\% \\

  \multicolumn{9}{l}{\textbf{Physical Disabilities}} \\
  Upper limb disability      & 12  & 1.1\%  & 4   & 33.3\% & 2  & 16.6\% & 6   & 50.0\% \\
  Lower limb disability       & 5   & 0.4\%  & 2   & 40.0\% & -- & --     & 3   & 60.0\% \\

  \multicolumn{9}{l}{\textbf{Visual Disabilities}} \\
  Blurred vision              & 7   & 0.6\%  & 1   & 14.2\% & 1  & 14.2\% & 5   & 71.4\% \\
  Color blindness             & 5   & 0.4\%  & 1   & 20.0\% & -- & --     & 4   & 80.0\% \\

    Myopia                      & 1   & 0.09\% & --  & --     & -- & --     & 1   & 100\% \\

  \multicolumn{9}{l}{\textbf{Others}} \\
  Social anxiety              & 3   & 0.2\%  & --  & --     & 1  & 33.3\% & 2   & 66.6\% \\
  Asthma                      & 1   & 0.09\% & 1  & 100.0\%     & -- & --     & --   & -- \\

  \bottomrule
  \end{tabular}
}
\end{center}

\end{landscape}
\clearpage

\subsection{RQ3. What is the categorization of accessibility reviews of VR applications?}

We identified the distribution of VR application categories for the top 20, the lowest-rated 40, and the most popular 10 applications, along with accessibility reviews associated with different types of disabilities. In Action VR applications,  accessibility reviews associated with motion sickness were more often positive (258; 51.0\%) than negative (165; 32.6\%), with 82 reviews (16.2\%) classified as neutral. Other disability-related accessibility issues were mentioned infrequently. Among these, negative feedback from users with epilepsy, and learning disability accounted for 50\% or more of the reviews. Conversely, positive feedback from users with autism spectrum disorder (ASD), upper limb disability, lower limb disability, hearing impairment, and color blindness was less than or equal to 50\%.

In Sports VR applications, motion sickness was the most frequently mentioned issue in accessibility reviews. Overall, reviews were more often positive (33; 42.3\%) than negative (27; 34.6\%), with 14 reviews (17.9\%) classified as neutral. All reviews associated with epilepsy (2; 100\%) provided negative feedback. Other disability-related accessibility issues were mentioned infrequently, including lower limb disability and ADHD.

In Social VR applications, accessibility reviews associated with motion sickness were more often negative (58; 72.5\%) than positive (10; 12.5\%), with 12 reviews (15.0\%) classified as neutral. Reviews associated with hearing impairments were also predominantly negative (114; 91.9\%), with 10 neutral reviews (8.0\%). For speech impairment, reviews were more negative (23; 60.5\%) than positive (2; 5.2\%), while 13 reviews (34.2\%) were neutral. Overall, hearing impairments and motion sickness were the most frequently mentioned disabilities in Social VR applications. Other disabilities, such as epilepsy, ASD, upper limb disability, blurred vision, social anxiety, myopia and color blindness, were also more often associated with negative feedback.

In Puzzle VR applications, accessibility reviews associated with motion sickness were more often positive (46; 55.4\%) than negative (15; 18.0\%), with 22 reviews (26.5\%) classified as neutral. Some disabilities, such as epilepsy, ASD, ADHD, and blurred vision, were only associated with negative feedback. Only two disabilities, upper limb disability and lower limb disability, were associated with positive feedback.

In Horror VR applications, only a few reviews mentioned accessibility issues, and most of these were negative. Only motion sickness received positive feedback. Hearing impairments accounted for the most frequently mentioned negative feedback.

In Simulation VR applications, accessibility reviews associated with motion sickness were more often positive (33; 47.8\%) than negative (20; 28.9\%), with 16 reviews (23.1\%) classified as neutral. Positive feedback was provided for three categories of disabilities: blurred vision, asthma, and cerebral palsy. Negative feedback was associated with two categories: ASD, and upper limb disabilities.

\begin{figure*}[t]
    \centering
    \includegraphics[width=\linewidth]{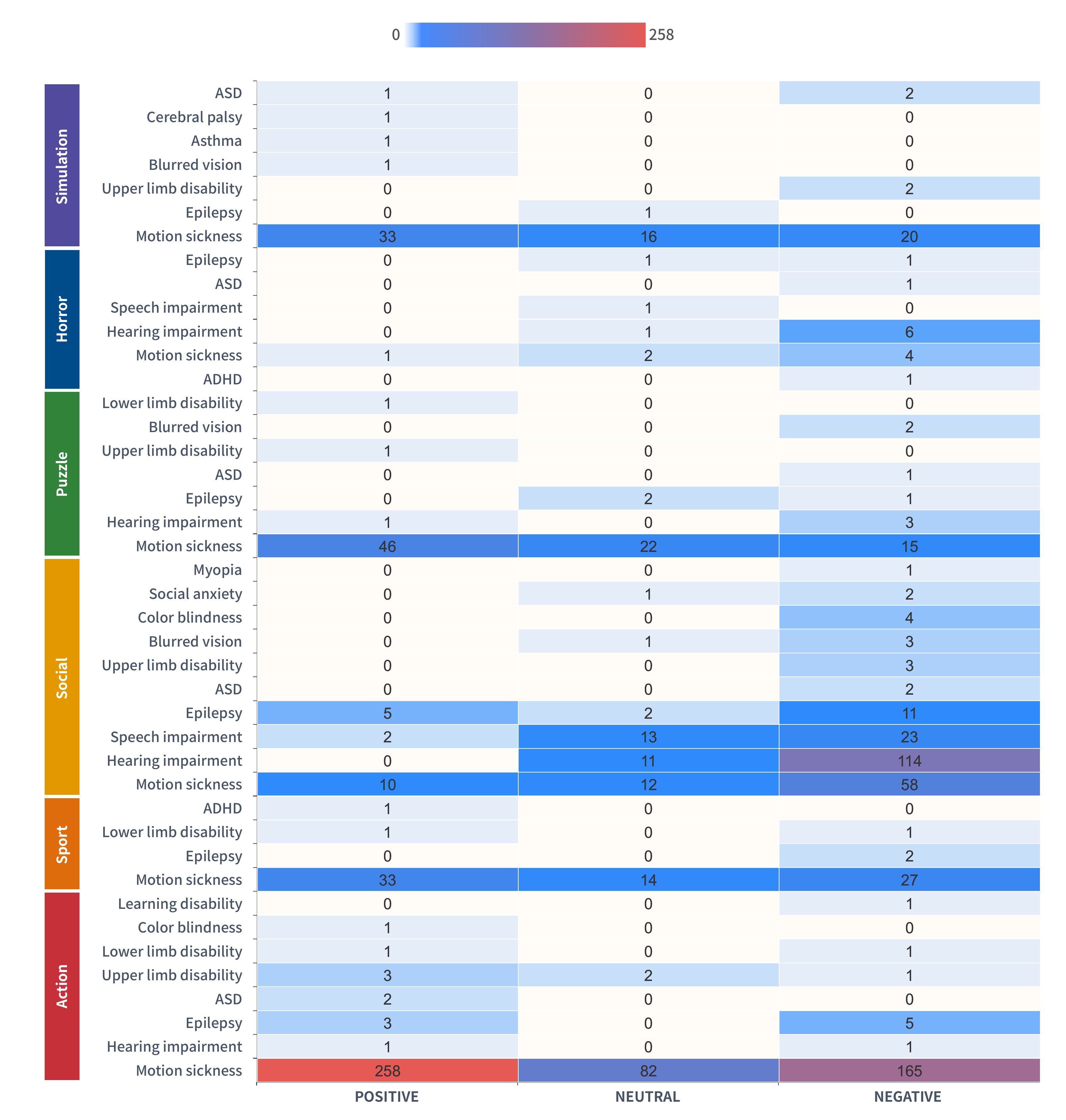}
    \caption{The information of the number of accessibility reviews associated with different types of disabilities and categorized VR applications.} 
    \label{fig:hot}
\end{figure*}

Figure \ref{fig:hot} shows the distribution of accessibility reviews associated with disabilities, depending on the categories of VR applications. Each cell indicates the number of accessibility reviews associated with disabilities for different types of VR applications. Motion sickness was the most common issue across all application types (\textit{N} = 505), except in Social VR applications, where hearing impairments were most prevalent. Similarly, speech impairments were also most common in Social VR applications. The number of accessibility reviews associated with other disabilities was low.

\begin{tcolorbox}[
  colback=white,        
  colframe=black,       
  arc=3mm,              
  boxrule=1pt,        
  left=5pt, right=5pt,  
  top=5pt, bottom=5pt,   
  breakable
]

\noindent
\textbf{RQ3 Findings: } 

\begin{itemize}
    \item There are significant differences in the distribution of accessibility reviews across different types of VR applications.
    \item Action VR applications received the highest number of accessibility reviews, with motion sickness accounting for the largest proportion of feedback (more positive than negative, alongside a certain share of ambiguous comments). Other types of impairments were mentioned far less frequently.
    \item Sports VR applications were dominated by motion sickness–related feedback, with positive comments slightly outnumbering negative ones.
    \item Social VR applications were predominantly concerned with hearing and speech impairments, with negative feedback comprising the overwhelming majority.
    \item In puzzle VR applications,  feedback related to motion sickness was generally positive, whereas feedback concerning other impairments, such as epilepsy, and ASD, was predominantly negative.
    \item Horror VR applications received very few accessibility-related comments, most of which were negative.
    \item In simulation VR applications, feedback related to motion sickness was slightly more positive than negative, and some positive experiences were also reported by users with ASD, visual impairments, lower limb disabilities, and cerebral palsy.
\end{itemize}

\end{tcolorbox}

\subsection{RQ4. What are the reasons underlying accessibility issues for users with disabilities?}




To answer RQ4, we identified three categories of underlying causes contributing to VR accessibility issues: user impairments, application design barriers, and user adaptations. We present user reviews organized by disability classification to highlight the specific factors contributing to accessibility challenges in VR. In this sub-section, \faIcon[regular]{thumbs-down} indicates negative reviews, \faIcon[regular]{hand-paper} indicates neutral reviews, and \faIcon[regular]{reply} indicates that the developer responded to the user’s reviews.

\textbf{1. Motion Sickness. }This was the most frequently mentioned issue in VR accessibility-related reviews. Its causes can be grouped into application design barriers, user adaptation, and user-specific impairments. Application design barriers include offering only smooth locomotion without a teleportation option, poorly designed control schemes, movement speeds or turning methods that induce discomfort, jumping or climbing actions, visual blurriness and low frame rates, visual quality differences across devices, and the absence of saveable or easily accessible comfort settings. Specifically, users reported issues related to object swaying, which the developers acknowledged and subsequently addressed.





\faIcon[regular]{thumbs-down} \textit{``...I did everything right and had a few turns with the lawnmower. Oh my god I was so sick.'' }

\faIcon[regular]{reply} \textit{``We’ve just pushed an update live that we hope will address this, with the addition of several new comfort settings. You’re now able to select from a variety of different vignette options and disable grass sway. We’ve also improved our centre dot feature.''}

Movement speed, turning visuals, and jumping in VR were identified as contributors to motion sickness.

\faIcon[regular]{thumbs-down} \textit{``...I did set turning style to snap turning too, because the default smooth turning was very uncomfortable to me...''}

\faIcon[regular]{thumbs-down} \textit{``the movement speed and style gave me a headache and motion sickness, and the touch pad on left controller to control movement felt to sensitive as well.''}

\faIcon[regular]{thumbs-down} \textit{``...But the movement causes insane Motion Sickness which I have not even experienced in VR rollercoastersims (especially when jumping).''}



Many users noted that first-person smooth locomotion (\textit{smooth movement}) tended to aggravate motion sickness during VR experiences. Conversely, some VR applications incorporated teleportation specifically to mitigate motion sickness.

\faIcon[regular]{thumbs-down} \textit{``No option for teleportation which isn’t ideal, it gave me a massive migraine and will have to sit in the dark today.'' }


\faIcon[regular]{thumbs-down} \textit{``The setting for selecting teleport won't save for me, so I have to manually put it from locomotion to teleport each time.'' }

Some users reported that the teleportation configuration was insufficiently discoverable, making it difficult to access or adjust during use.

\faIcon[regular]{thumbs-down} \textit{``I mean in like 10 minutes.There is no teleportation locomotion implemented, so you have to move and rotate by controller / thumb input, which leads to sickness feelings.''}

Conversely, visual related issues contributed to motion sickness. For example, a lack of visual detail and image blurriness were reported as contributing factors.

\faIcon[regular]{thumbs-down} \textit{``...I actually have a lot of fun with this game, however it’s SO blurry and unsmooth with terrible textures, when seeing stuff like that in VR it’s not only not immersive at all, but it also causes motion sickness and eye strain MUCH easier and quicker than other games, PLEASE OPTIMIZE, it’s not the quest 2’s fault.''}



Some users also reported that variations in visual quality across different devices, as well as the disabling of the open-source tool OpenFSR (Open FidelityFX Super Resolution), contributed to motion sickness.

\faIcon[regular]{thumbs-down} \textit{``...I noticed some stutter in the graphics, which might be the reason for discomfort. This was on a Quest 2.''}

\faIcon[regular]{thumbs-down} \textit{``...Let alone things like OpenFSR which was critical in improving performance, which helped reduce motion sickness for those who may not be acclimated to VR.''}

Some motion sickness issues were identified in social VR applications.

\faIcon[regular]{thumbs-down} \textit{``...Game not suited for motion sickness people, most players shun the teleport feature enough to vote kick/ban you (more on that later) and most people who create rooms outright remove said feature, so, to say the least it isn’t friendly for motion sickness people.''}

Some users experienced motion sickness while using the VR application in a standing position. Developers recommended using the application in a seated position to mitigate this issue.

\faIcon[regular]{thumbs-down} \textit{``The game is fun to play but not for more than 30 minutes... Then I'm so dizzy and I feel like I want to throw up, this might not be the case for everyone though.''}

\faIcon[regular]{reply} \textit{``If you're experience motion sickness we recommend playing seated, it should help you out.''}


We also found that some users reported improvements in motion sickness symptoms after extended use. These comments were most often found in neutral reviews.

\faIcon[regular]{hand-paper} \textit{``I am big into the [App name] and I used to get motion sick from turning and sudden stops. Now, after about 2 weeks, I barely notice it. You just need time to adjust.''}

Another example revealed that comfort settings had limited impact, with improvement in motion sickness primarily attributed to prolonged use of VR applications.


Finally, some elderly users also reported a reduction in motion sickness symptoms with continued use.

\faIcon[regular]{hand-paper} \textit{``I'm 75, never played computer games before... At first I couldn't stand more than 10 minutes without feeling queasy, but now I can enjoy a full session. VR has really opened a door for me.''}

\textbf{2. Hearing Impairments. }This was a common accessibility issue identified in user reviews concerning VR accessibility. Reported issues included the lack of subtitle and textual information support, insufficient auditory cues, limited hearing-friendly features in social applications, and accessibility challenges in multi-user interactions. Regarding the lack of subtitle and textual information support, some users noted that the absence of subtitles negatively affected their sense of engagement during gameplay.

\faIcon[regular]{thumbs-down} \textit{``...One thing that disappoints me is that I’m deaf, so I can’t hear when the...walks, makes loud noises, or does anything similar. For hearing players, that’s perfect—but not for me...''}

Most reviews from users with hearing impairments originated from social and multi-user VR applications. Users reported that successive updates to these platforms had removed several accessibility features.

\faIcon[regular]{thumbs-down} \textit{``Modding added nice quality of life changes... and added features to help those with disabilities (such as text to speech and caption support for the hard of hearing)...While modding has always been against the TOS of [App name], actively going after it and disallowing it outright hurts the user experience majorly...[App name] claims they're going to add some of these features... but... they will never be able to keep up with the numerous features users want or need.''}

\faIcon[regular]{thumbs-down} \textit{``I was excited because I'm a huge fan of paranormal stuff...but before I can start to play... I clicked on options to turn the subtitles on but there is none...I am deaf/hard of hearing and I feel this isn't fair.''}

Some users attributed this issue to the developers’ introduction of the Easy Anti-Cheat (EAC) system.

\faIcon[regular]{thumbs-down} \textit{``Modding added nice quality of life changes... and added features to help those with disabilities (such as text to speech and caption support for the hard of hearing)...While modding has always been against the TOS of [App name], actively going after it and disallowing it outright hurts the user experience majorly...[App name] claims they're going to add some of these features... but... they will never be able to keep up with the numerous features users want or need.''}

\faIcon[regular]{thumbs-down} \textit{``Recently they have implemented a very wrong decision of adding EAC. It has removed a certain level of safety to the community and accessibility...the accessibility is for those that may be hard of hearing or cannot hear all together...''}

Some users with disabilities reported that certain VR applications lacked accessibility support, with community developers openly stating that they had no plans to incorporate such features in future updates, in some instances, even expressing discriminatory attitudes toward these users.

\faIcon[regular]{thumbs-down} \textit{``However, it's become clear that their developers don't care about people with disabilities. [Developer's name] stated on Discord that they will not implement certain accessibility features because, while they would make the game playable for people with disabilities, they would also make it easier for non-disabled players. They also joked about deaf and hard of hearing gamers and condoned community members mocking blind players.''}

\textbf{3. Speech Impairments.} Users with speech impairments commonly reported significant limitations in self-expression within VR environments. To address this, many recommended the inclusion of non-verbal communication mechanisms and text chat functionalities as essential accessibility features.

\faIcon[regular]{thumbs-down} \textit{``...say that the lack of proper text chat makes this game unplayable for mute users like me.''}

\faIcon[regular]{thumbs-down} \textit{``As a mute user, non-verbal presence is the only way I can exist in these virtual spaces.''}

One user noted that avatar movements could serve as a form of non-verbal communication; however, this approach had not been well developed or effectively implemented.

\faIcon[regular]{thumbs-down} \textit{``{App name} had potential for users who have mute/DHH/deaf/autistic issues to express themselves through avatars and actions, but it got worse over time.''}

Moreover, some users with speech impairments reported experiencing social exclusion within certain VR environments, highlighting the lack of inclusive design in social interactions.

\faIcon[regular]{thumbs-down} \textit{``Soo... it's a bit of a hot mess right now. [App name] used to be inclusive, but now people who don’t use voice chat are often ignored or kicked.''}

\textbf{4. Epilepsy.} Users with epilepsy reported that certain visual characteristics, such as flickering images, intense lighting effects, and strong rhythmic patterns, were primary factors contributing to seizure episodes in VR environments.

\faIcon[regular]{thumbs-down} \textit{``After booting up the game today I noticed that some of the new lighting effects are extremely bright and could trigger seizures.''}

Some examples highlighted that certain VR applications lacked essential accessibility configurations, such as warning systems for potentially harmful visual stimuli and user settings to disable flashing effects.

\faIcon[regular]{thumbs-down} \textit{``There are no warnings about flashing lights. As someone with epilepsy, I find this really irresponsible.''}

\faIcon[regular]{thumbs-down} \textit{``I've changed every setting in gameplay/graphics and still can't disable the flashing effects.''}

In addition, some users with epilepsy reported that, in social VR applications, other users could maliciously employ flashing lights.

\faIcon[regular]{thumbs-down} \textit{``...them are crasher or just lots of flashing lights do not suggest for people that have Epilepsy.''}

\textbf{5. Physical Impairments.}Most users with physical disabilities reported impairments in either their left or right hand and expected VR applications to support single-handed interaction. One user with a right-hand disability noted that the VR application required two-handed interaction, making it difficult to continue using.

\faIcon[regular]{thumbs-down} \textit{``I am disabled and can use only my right hand to play the game, using the vive controller. The game works well for me until I am to get the [game object]...And now I have to use two hands. I am unable to play the game any further.''}

Additionally, during data processing, we found that some users without physical disabilities also expressed a need for specific accessibility features (these reviews were not included in the final dataset). For example, left-handed users requested support for left-hand interactions, while pregnant users reported difficulty standing for extended periods and therefore preferred using VR applications in a seated position.

\textbf{6. Other Impairments.} For visual impairments, most negative reviews concerned the disabling of modding in social VR applications. Similarly, for users with autism, some negative reviews was directed at social VR applications, with concerns that exposure to negative social content could adversely affect their well-being. Some users with autism also raised concerns about the accessibility of certain VR applications. For example, one user expressed a desire for the addition of a single-player mode.

\faIcon[regular]{thumbs-down} \textit{``People who love the social side of the game can continue to enjoy the co-op modes, but, people like me, who sometimes prefer to be lone wolves and have really debilitating phases of social anxiety, can enjoy the game in their own way.''}

For users with learning disabilities, one user noted that the multiplayer mode (the forced participation in competitive play) restricted their gaming experience.

\faIcon[regular]{thumbs-down} \textit{``...Because we'll as much I hate to admit and am angry about how bad I am...and for the people judging me for that. I have a Learning disability.''}

\begin{tcolorbox}[
  colback=white,        
  colframe=black,       
  arc=3mm,              
  boxrule=1pt,        
  left=5pt, right=5pt,  
  top=5pt, bottom=5pt,   
  breakable
]

\noindent
\textbf{RQ4 Findings: } 

\textbf{Motion Sickness: }
\begin{itemize}
    \item The causes included: offering only smooth locomotion without a teleportation option, poorly designed control schemes, visually uncomfortable movement speed or turning, jumping and climbing actions, visual blurriness and low frame rates, visual quality differences across devices, and the absence of saveable or easily accessible comfort settings.
    \item Some users reported that prolonged use gradually alleviated their symptoms, although for others it remained a significant barrier.
\end{itemize}

\textbf{Hearing Impairments: }
\begin{itemize}
    \item Common issues included the absence of subtitles or textual information, a lack of hearing-friendly features (e.g., volume amplification, adjustable audio distance), and the absence of alternative communication methods in multiplayer interactions.
    \item This issue was particularly pronounced in social VR, where certain updates, such as the introduction of anti-cheat systems, effectively disabled third-party accessibility plugins, resulting in a regression in accessibility.
\end{itemize}

\textbf{Speech Impairments: }
\begin{itemize}
    \item The absence of text chat and nonverbal communication mechanisms led to the marginalization of mute users.
    \item It was recommended to incorporate nonverbal interaction designs.
\end{itemize}

\textbf{Epilepsy: }
\begin{itemize}
    \item Triggers included flashing visuals, intense light effects, and the absence of photosensitivity warnings or options to disable flashing effects.
    \item In social VR, there was also a risk of malicious use of flashing elements to carry out attacks.
\end{itemize}

\textbf{Physical Impairments: }
\begin{itemize}
    \item Users with upper limb disabilities faced difficulties in performing mandatory two-handed operations, while those with lower limb impairments experienced limitations in applications that only support standing interactions.
    \item There was a need for a single-hand mode (supporting both left- and right-handed use), seated/lying interaction support, and alternative gesture-based operation options.
\end{itemize}

\textbf{Visual Impairments: }
\begin{itemize}
    \item Negative reviews were largely concentrated on social VR applications that, after disabling mods (third-party modification tools), prevented users from adjusting elements such as text size, color, brightness, and contrast, thereby removing essential visual accessibility support.
\end{itemize}

\textbf{Cognitive impairments: }
\begin{itemize}
    \item Users with cognitive impairments, such as autism, social anxiety, and learning disabilities, tended to prefer low-social-stimulation or single-player modes.
\end{itemize}

\end{tcolorbox}

\section{Discussion}\label{dis}

In this section, we discuss the results derived from accessibility reviews of the top 40, the most popular 20, and the lowest-rated 40 VR applications on the Meta and Steam Stores.

\subsection{Distribution of Accessibility Reviews}

In our sample, reviews reporting accessibility issues affecting users with disabilities were scarce. Aside from motion sickness, hearing impairments, speech impairments, epilepsy, and upper limb disabilities, the counts for all other disabilities were in the single digits (\textit{N $<$ 10}). Unexpectedly, the top-rated and most popular Horror VR applications on the Meta Store had no accessibility reviews. Among the 40 lowest-rated VR applications, only 46 accessibility reviews were identified. The scarcity of accessibility reviews might be attributed to three main reasons: 1. We did not select user reviews from other VR application stores, such as PlayStation VR\footnote{https://www.playstation.com/en-us/ps-vr/} and Pico\footnote{https://www.picoxr.com/global}, because our preliminary investigation indicated that these VR platforms had less diversity in applications and smaller user bases compared to Meta and Steam. 2. Users with disabilities may be reluctant to provide accessibility reviews, possibly due to concerns about discrimination. It is also possible that they have reported accessibility issues to developers via Discord\footnote{ (https://discord.com/}, as some VR applications maintain their own communities on Discord or other platforms. 3. The number of users for the lowest-rated VR applications was significantly lower than that of the top and most popular VR applications.






Accessibility reviews were more negative than positive. Prior research similarly found that most users with disabilities tend to provide negative reviews \cite{kraft2001customer}. On the Steam store, negative reviews were more common in Social VR applications, whereas positive reviews were more prevalent in Puzzle VR applications. On the Meta store, Simulation VR applications more frequently received positive reviews. Notably, Action and Social VR applications accounted for a higher volume of accessibility reviews overall. Within these categories, positive reviews were more common in Action VR applications, while negative reviews were more prevalent in Social VR applications. Additionally, neutral reviews indicated that although some users with disabilities encountered accessibility challenges, they still considered using the VR applications due to their engaging narratives and content.


Furthermore, even fully developed mobile applications still exhibited a small number of accessibility barriers \cite{Yan19}. Compared to mobile applications, VR applications might encounter even greater accessibility challenges \cite{Naikar24}. Specifically, the XAUR was at the draft stage, which indicates that comprehensive VR accessibility guidelines for practitioners are currently lacking. Although various VR companies have their own defined accessibility guidelines, such as those from Meta. Their guidelines may not be applicable to other VR headsets. Therefore, incorporating accessibility considerations when releasing a VR application on platforms such as Steam and Meta could impose financial or time pressures on enterprises of varying sizes, particularly startups. As a previous study found, small IT companies or startups rarely consider accessibility requirements \cite{WANG2025107609, Wang25Understanding}. Furthermore, a lack of reviews did not imply either satisfaction or dissatisfaction with VR applications, as some users who encounter frustrations are unlikely to share their feedback \cite{hydock2020unhappy}. Compared to previous studies on mobile app accessibility reviews, the number of accessibility reviews we collected for VR applications was significantly lower than that for mobile Apps \cite{Oliveira23, Aljedaani22, Eler19}. During the data analysis phase, We found that discussing accessibility requirements through community channels is also a viable approach. However, some users with disabilities reported that community developers were unwilling to address accessibility issues if doing so might affect the experiences of general users. These findings highlight the need for future research to explore the attitudes of VR community developers and users with disabilities toward accessibility.


\subsection{Causes of Accessibility Issues}

A total of 17 types of disabilities were identified in our study. Among these, motion sickness was the most commonly mentioned accessibility issues across various VR application categories, participially in action VR applications. Motion sickness was the only accessibility reviews where positive outweighs negative reviews. It also accounted for almost half of all accessibility reviews. A previous study identified hardware, content, and human factors as the primary causes of motion sickness in VR \cite{chang2020virtual}. However, no human factors were mentioned as contributing causes. Instead, all motion sickness reviews were associated with hardware setups and VR content. Many of these causes align with previous studies; however, we also found that some users with disabilities reported rare and unique causes of motion sickness. 

\begin{itemize}
    \item Sound effects can cause motion sickness in some users with accessibility issues. Previous studies indicated that motion sickness might be caused by purely auditory stimuli in front of a curved projection display with a visual scene \cite{keshavarz14}. However, it remains unclear whether sound effects are a direct factor in VR. For example, we are unsure whether noisy sound effects are a primary or secondary influencing factor. 
    \item Swaying UIs may contribute to discomfort. Some users prone to motion sickness reported that swaying interfaces made them feel uneasy, and reviews also indicated that the swaying objects could induce motion sickness. However, it remains unclear whether swaying UIs are indeed one of the causes of motion sickness. Additionally, adaptive UIs should be considered in various VR application \cite{Gajos08, Lu24}.
    \item Extended use of a VR application can help some users with mild or moderate motion sickness alleviate their symptoms \cite{zhang2025long}. This finding motivates further research into how demographic factors (such as age and gender) and content, interaction, or hardware factors can mitigate motion sickness. In addition, we found that older users were also able to reduce motion sickness. However, it remains unclear whether the adaptation period differs across age groups.
   
\end{itemize}


Hearing and speech impairments were another major focus of accessibility reviews in social VR applications. This may be due to the reliance of social VR on communication and collaboration among diverse users. Specifically, many users with hearing impairments noted that VR applications lack subtitles. Some VR applications might have passive voice subtitles, such as those for NPCs (Non-player characters) with constructed closed captions \cite{Coutinho11}, but real time communication rarely provided subtitles. Meanwhile, some users noted that the subtitle feature is disabled when VR application (social) updated their anti-cheat systems. A previous study also indicated that software updates could result in incompatibilities with accessibility features \cite{DS24}. Some users with hearing impairments reported that volume enhancement could help them hear sounds clearly in virtual environments. However, this is not considered a critical feature, as hearing aids can assist users with hearing impairments. Furthermore, users with speech impairments reported difficulties in real-time communicate with others, highlighting the essential role of text-to-speech feature. Despite the availability of text input capabilities in several popular social VR applications, these messages were often disregarded, leading users with speech impairments to feel isolated in communication.

Sign language was a special accessibility feature in VR that received positive reviews from users with hearing and speech impairments. Sign language not only facilitates communication but also enriches understanding of other countries' cultures. However, its use was predominantly confined to VR social applications, such as VRChat\footnote{https://hello.vrchat.com/}, and RecRoom\footnote{https://recroom.com/}. Furthermore, sign language might be the most effective method of communication for deaf and mute users. Other primary factors for sound accessibility included voice isolation, volume control, and audio distance. These three factors were primarily mentioned in the context of social VR applications. Among these, some users with hearing impairments considered that audio distance causes difficulties in discerning sounds. A previous study indicated that distance is an indispensable cue for audio in VR \cite{McArthur17}. However, as some users noted the VR community acknowledges that accessibility features may influence the experience of general users. Nonetheless, features such as controllable audio distance and enhanced audio feedback  significantly benefit individuals with hearing impairments. Previous studies also indicated that while accessibility design and features may address the needs of specific user groups, they can sometimes lead to operational discomfort for general users \cite{heilemann2021accessibility, creed2024inclusive, chong2021virtual}. Thus, achieving a balance between usability and accessibility (also known as usability and accessibility win-win in Universal Design \cite{goldsmith2007universal}) remains a key research direction within the VR and human-computer interaction communities. Meanwhile, it is necessary to conduct a comprehensive investigation into whether the VR accessibility features of existing VR applications meet accessibility standards.

Epilepsy was frequently mentioned as a concern by users with disabilities. Photosensitivity was one of the main symptoms of epilepsy. Many VR applications had general warnings for users with epilepsy, yet these users often remain excluded from VR experiences \cite{tychsen2020concern}. Some users with epilepsy reported that they are susceptible to malicious attacks in social VR applications such as VRChat and RecRoom, particularly through the use of flashing 3D avatars and images. This finding aligns with a recent empirical study and identifies reasons \cite{South24}.

Four types of cognitive impairments were identified as prevalent in the reviews: autism spectrum disorder (ASD), attention deficit hyperactivity disorder (ADHD), social anxiety disorder (SAD), and learning disability. Among these, users with ASD, SAD and learning disability mentioned the single-player mode. They indicated that this mode can reduce their stress, as some users may face criticism during collaborative activities. However, most VR applications, except for standalone VR applications, were noted to lack a single-player mode. Meanwhile, sound effects can intensify their negative emotions. Previous studies also indicated that while negative or frightening sound effects can intensify anxiety in users with autism, integrating exposure therapy techniques into VR games may alleviate perceived anxiety due to auditory stimuli \cite{johnston2020soundfields, Maskey2014ReducingSP}. Additionally, user with social phobia mentioned that communicating with other users may cause them anxiety. Users with ADHD noted that the abundance of diverse 3D avatars, noisy conversations, and rich interaction methods in social VR makes it difficult for them to maintain focus. Meanwhile, complex interaction methods could lead to difficulties in maintaining focus. Additionally, engaging story lines helped users with ADHD maintain focus, while dull story lines had the opposite effect. Overall, although many studies have evaluated that VR can help mitigate the effects of cognitive impairments, challenges persist in existing VR applications for those with cognitive disabilities. These issues may be more pronounced in lower-rated VR applications. As the XAUR guidelines suggest, VR applications should personalize the immersive experiences in various forms to better assist users with cognitive and learning disabilities. 

Among the visual impairments discussed, only blurred vision, myopia and color blindness were mentioned. Specifically, one user disclosed suffering from myopia. Specifically, some VR applications allowed users to position text closer to their avatars for easier reading. Meanwhile, text size and color were key factors affecting users with blurred vision. A previous study found that accessibility reviews collected for mobile applications also mention text size and color \cite{Oliveira23}. However, text color in virtual environments might be influenced by additional factors, such as canvas color, background color, and ambient lighting. As the XUAR guidelines suggest, VR applications should allow user customization of components and resources. Additionally, low brightness can cause users with blurred vision to struggle with seeing both 3D and 2D content clearly in virtual environments. A previous study developed a toolkit (\textit{SeeingVR}) that supports the adjustment of brightness, contrast, and recoloring of 3D objects in virtual environments \cite{Zhao19}. This not only enables users with low vision to see clearly but also benefits users with epilepsy by moderating brightness and contrast~\cite{South24}.

There were two types of physical disabilities, including upper limbs (hands or arms) and lower limb. Users with physical disabilities mentioned that two common issues in accessibility reviews are the fast tempo, which can lead to a poorer experience for them and the inability to respond promptly in virtual environments. Furthermore, many users with upper limbs disabilities reported a poor experience when using VR with one-handed operation. During the data analysis phase, we also found that left-handed users also reported a poorer experience (excl. in final dataset). Although we did not identify any left-handed users with upper limb disabilities in the accessibility reviews, we recommend that one-handed modes should consider including options for both left-handed and right-handed users. Meanwhile, some users with upper limb disabilities reported difficulties in operating controllers and suggested that VR applications should incorporate gesture interactions. Some users with lower limb disabilities mentioned that many VR applications are only standing interactions, making it difficult to interact while lying down or sitting. This also can affect the usability for particular challenged user groups, such as pregnant women or the elderly. 

Other types of disabilities, such as asthma and cerebral palsy, were rarely mentioned in accessibility reviews. We noticed that a user with asthma benefits from painting simulation VR, which can prevent the inhalation of gases produced by paints. An user with cerebral palsy noted that VR application (Simulation) did not cause motion sickness but suggested simplifying the interactions.

Overall, motion sickness appears to be the most concerning accessibility issue across VR platforms. Many accessibility reviews only briefly mentioned common disabilities. For instance, some reviews refer simply to myopia when discussing blurred vision. Notably, we compared the XAUR guidelines with our findings and found that many of the underlying reasons were not included in the XAUR guidelines. Therefore, we recommend that future work considered the following: 1. The VR community should encourage users with disabilities to actively share their experiences with VR applications. 2. VR practitioners must place greater emphasis on accessibility reviews provided by users with disabilities to foster a more inclusive environment. 3. Both practitioners and researchers should collaborate to develop a more comprehensive and detailed set of XAUR guidelines.









\section{Threats to Validity}\label{limi}

This section presents several limitations of our study, including sampling bias and external validity risks.

\textbf{Sampling Bias. }Each step of the sampling process may introduce bias to our study. Some selected VR applications may not be used or mentioned by users with disabilities. To mitigate this threat, we selected the top and the most poplar VR applications from the Meta and Steam stores. We also aimed to understand the state of accessibility reviews from lowest-rated VR applications. Further, to ensure we identified most accurate and relevant accessibility reviews, we defined keywords corresponding to disability categories based on the World Health Organization's definitions of disabilities. We categorized these keywords into six types of disability as defined by XAUR. we also defined additional keywords based on the XAUR guideline, to ensure the identification of more accessibility reviews, such as voice command, time limited, navigation, and rest. Additionally, to avoid incorrect classification of accessibility reviews by researchers, we used cross-validation among researchers to ensure all parts involved agreed upon the decisions, such as inclusion, exclusion, and coding.

\textbf{External Validity. }We extracted a large dataset of user reviews from the Meta and Steam stores. We excluded some applications, such as mixed reality applications, or desktop applications on Steam stores, ensuring that only VR applications were included. In addition, we did not limit the selection of accessibility reviews because VR end-users are fewer in number than mobile end-users. Choosing specific accessibility reviews could result in insufficient data and a lack of comprehensive investigations. Finally, we could obtain a richer set of accessibility reviews if we included other VR stores (e.g., PlayStation VR) and reviews in different languages from various countries.

\section{Conclusion}\label{con}

This study investigates VR accessibility reviews posted by users with disabilities. Our dataset comprised 1,367,419 reviews collected from the top 40, the 20 most popular, and the 40 lowest-rated VR applications on the Meta and Steam Stores. From this dataset, we extracted 1,076 accessibility-related reviews. To the best of our knowledge, this is the first investigation of VR accessibility reviews. Our results highlighted that accessibility reviews were more frequently negative than positive. Motion sickness and hearing impairments were the most commonly reported issues. Motion sickness was most frequently mentioned in Action VR applications, while hearing and speech impairments were most often referenced in Social VR applications. Accessibility issues associated with other disabilities were mentioned less frequently. Users with motion sickness and hearing impairments provided the most diverse range of positive and negative feedback regarding VR accessibility. In contrast, some disabilities raised only a single concern each, such as learning disabilities. In addition, we analyzed the causes of accessibility issues based on negative and neutral VR accessibility-related reviews.

\begin{acks}
Grundy is supported by ARC Laureate Fellowship FL190100035.
\end{acks}

\bibliographystyle{ACM-Reference-Format}
\bibliography{ref}

\newpage

\appendix

\section{Appendix}

\subsection{Keywords \& Names of VR Applications}

\begin{table}[h]
\caption{XAUR-aligned disability categories and associated keywords.}
\label{tab:vr-acc-keywords-by-theme}
\small
\setlength{\tabcolsep}{4pt}
\begin{tabular}{p{0.30\linewidth} p{0.66\linewidth}}
\hline
\textbf{Theme} & \textbf{Keywords} \\
\hline
Immersive semantics and customization & Immersive semantics; Customization; Assistive technology; Navigation; Identify locations; Identify; Interaction; Immersive environment; Intuitive; Location descriptions; Object descriptions; Repositioning; Resizing; Sensitivity adjustment; Contextual; Filter; Sort; Query \\

Motion-agnostic interactions & Motion-agnostic interactions; Interact with items; Bodily movement; Device-independent; Perform actions; UI access; Input; Multiple input \\

Immersive personalization & Immersive personalization; Personalize experience; Symbol sets; Communication; Overlays; Affordances; User preferences; Mute non-critical content; Turn off animations; Visual content; Audio content; Non-critical messaging \\

Interaction and target customization & Interaction customization; Target customization; Target size; Button size; Fine motion control; Activate input; Hit targets; Target spacing; Multiple actions; Multiple gestures; Sticky keys; Serialization; Press multiple buttons \\

Voice commands & Voice commands; Voice activation; Navigate; Interact; Communicate; Voice control; Screen readers; Voice assistants; Device pairing; Eliminate extra steps \\

Color changes & Color changes; Customize colors; Indicate danger; Indicate permission; High-contrast themes; Luminance; Color-contrast \\

Magnification context and resetting & Magnification context; Reset; Screen magnification; Check context; Track focus; Reset focus; Interface elements; Enlarge elements; Menu reflow; Usability \\

Critical messaging and alerts & Critical messaging; Alerts; Focus; Second screen; Priority roles; Flags \\

Gestural interfaces and interactions & Gestural interfaces; Virtual menu system; Touchscreen accessibility; Accessibility gestures; Swipes; Flicks; Taps; Self-voicing; Spoken descriptions; Gesture input; Query items; Gesture remapping; Virtual switch; Macros; Change defaults\\

Signing videos and text transformation & Signing videos; Text description transformation; Signing avatar; Video for text; Object/item descriptions; Pre-recorded signing; Signer size; Video stream size; Sign language \\

Safe harbour controls & Safe-harbour controls; Safe place; Hotkey; Shortcut; Manually defined \\
Immersive time limits & Time limits; Digital wellbeing; Lose track of time; Platform integration; Alarms; Session control \\

Orientation and navigation & Orientation and navigation; Maintain focus; Reset/calibrate orientation; Device-independent; Field of view; Personalized view; Visual/audio landmarks \\

Second screen devices & Second-screen devices; Real-time communication; Routing requirements; Text output; Alerts; Environmental sounds; Audio routing; Critical messaging; Content display; Navigate menus \\

Interaction speed & Interaction speed; Change speed; Perform interactions; Modify/extend timings; Critical inputs; User help; Start/stop mechanisms \\

\hline
\end{tabular}
\end{table}

\begin{table}[h]
\caption{XAUR-aligned disability categories and associated keywords.}
\label{tab:vr-acc-keywords-by-theme2}
\small
\setlength{\tabcolsep}{4pt}
\begin{tabular}{p{0.30\linewidth} p{0.66\linewidth}}
\hline
\textbf{Theme} & \textbf{Keywords} \\
\hline

Avoiding sickness triggers & Motion sickness; Teleportation movement; XR interactions; Smooth motion; Flickering images; Seizure-prevention settings; Reduce/turn off flicker \\

Spatial audio tracks and alternatives & Spatial audio; Audio accommodations; Spatialized audio; 3D sound; Text descriptions \\

Mono audio option for spatial orientation & Spatial orientation; Mono-audio option; Stereo/binaural soundscape; Mono audio; Perceive soundscape \\

Captioning, subtitling and text & Captioning; Subtitling; Multimedia content; Customizable captions; Text content \\

Immersive environment challenges & Input devices; Control; Timing and simultaneous action;   \\

Various input modalities & Speech input; Voice commands; Keyboard input; Switch input; Gesture-based controllers; Eye tracking; Selection \\

Various output modalities & Tactile; Haptics; Visual; 2D/3D graphics; Auditory; Spatial audio; Surround sound; Olfactory; Gustatory; Feedback \\

XR controller challenges & Controller usability; Sensory substitution devices; Keyboard remapping; Simulated actions; Walk in VR; Motion range; Position and orientation \\

Customization of control inputs & Modify input preferences; Variety of input devices; Control movement \\

Consistent tracking with multiple inputs & Tracking; Switching input devices; Loss of focus; Unwanted inputs/choices; Synchronized outputs; Focus modification\\

Usability and affordances in XR & Usability; Affordances; Multiple modalities; Modality translation; Modality preferences; Context of use; Allowed/disallowed interactions; Modality synchronization; Caption synchronization; Real-time text transcription\\
\hline
\end{tabular}
\end{table}


\begin{table}[ht]
\small
\caption{WHO/ICF-aligned disability categories and associated keywords.}
\label{tab:who-disability-keywords}
\centering
\begin{tabular}{p{3.4cm} p{0.7\linewidth}}
\toprule
\textbf{Disability category} & \textbf{WHO/ICF-aligned keywords} \\
\midrule
\textbf{Auditory disabilities} &
hearing loss; hearing impairment; deaf; hard of hearing; auditory disability; tinnitus; auditory perception disorder; unilateral hearing loss; bilateral hearing loss; central auditory processing disorder \\

\textbf{Cognitive disabilities} &
attention deficit hyperactivity disorder (ADHD); attention deficit; learning disability; cognitive impairment; mild cognitive impairment; memory impairment; problem-solving impairment; intellectual disability; dementia; developmental delay; executive function disorder; information-processing difficulty; mental fatigue \\

\textbf{Neurological disabilities} &
autism spectrum disorder (ASD); cerebral palsy; Parkinson’s disease; multiple sclerosis; epilepsy; seizure disorder; traumatic brain injury (TBI); stroke-related impairment; muscular dystrophy; Guillain–Barré syndrome; neuropathy; motor neuron disease; ataxia; motion sickness; cybersickness; neurodegenerative disorder \\

\textbf{Physical disabilities} &
upper-limb disability; lower-limb disability; mobility impairment; fine-motor impairment; wheelchair user; amputation; arthritis; osteoporosis; limb deformity; muscular weakness; spinal cord injury; postural instability; gait disorder; chronic pain; balance disorder \\

\textbf{Speech disabilities} &
speech impairment; speech disability; stuttering; dysarthria; aphasia; mute; non-verbal; voice disorder; laryngectomy; selective mutism; apraxia of speech; slurred speech; speech delay \\

\textbf{Visual disabilities} &
low vision; blurred vision; visual impairment; colour vision deficiency (CVD); color blindness; myopia; hyperopia; astigmatism; cataract; glaucoma; macular degeneration; retinitis pigmentosa; diabetic retinopathy; blindness; photosensitivity; hemianopia; tunnel vision \\
\bottomrule
\end{tabular}
\end{table}

\begin{table}[h]
\small
\caption{Steam Store: Names of VR Applications.}
\label{tab:who-disability-keywords}
\centering
\begin{tabular}{p{4.1cm} p{4.1cm} p{4.1cm}}
\toprule
\textbf{Top} & \textbf{Popular} & \textbf{Lowest-rated} \\
\midrule

  Car Mechanic Simulator  & Assetto Corsa  & Age of Heroes (VR) \\

  CarX Drift Racing Online &  Beat Saber & Block'hood VR\\

  Dagon by H. P. Lovecraft & Blade and Sorcery & Climb VR New York Parkour\\

  FIVE NIGHTS AT FREDDY'S HELP WANTED & Ghosts of Tabor & DrumMasterVR\\

  Half-Life Alyx & Gorilla Tag & Employee Recycling Center\\

  Keep Talking and Nobody Explodes & Hot Dogs, Horseshoes \& Hand Grenades & Endless Crusade\\

  Phasmophobia & Pavlov & Fishing Adventure VR\\

  Ragnarock & Rec Room & House Builder VR\\

  Resident Evil 7 Biohazard & The Elder Scrolls V Skyrim VR & Lucky Night Poker Games\\

  Tetris Effect Connected & VRChat & MADE VR  Interactive Movie - 01. Run away!\\

  Resident Evil Village && Relaxing VR Games Mahjong\\

  The Lab && Salvage Op\\

  The Room VR A Dark Matter && Seduction \\

  Universe Sandbox && SimpleProject\\

  Vermillion - VR Painting && The Peak Climb VR\\

  Vertigo 2 && VR Plane Crash\\

  VTOL VR && VR Prison Escape\\

  Walkabout Mini Golf VR && VR Shooting Range Multiple Weapons\\ 

  Subnautica && War Platform\\

  Spice \& Wolf VR2 && Zuma Legend VR\\

\bottomrule
\end{tabular}
\end{table}

\begin{table}[t]
\small
\caption{Meta Store: Names of VR Applications.}
\label{tab:who-disability-keywords}
\centering
\begin{tabular}{p{4.1cm} p{4.1cm} p{4.1cm}}
\toprule
\textbf{Top} & \textbf{Popular} & \textbf{Lowest-rated} \\
\midrule

  BIG BALLERS VR  & Beat Saber  & Behind The Frame The Finest Scenery VR \\

  Blacktop Hoops &  Blade \& Sorcery Nomad & Car Detailing Simulator\\

  Cubism & BONELAB & Car Mechanic Simulator\\

  Five Nights at Freddy's Help Wanted 2 & Gorilla Tag & Glue\\

  GOLF+ & Job Simulator & GRID Legends\\

  GYM CLASS - BASKETBALL VR & Penguin Paradise & Hatsune Miku VR\\

  HUMANITY & POPULATION ONE & Lawn Mowing Simulator VR\\

  Kartoffl & Rec Room & Microsoft Mesh\\

  Max Mustard & Roblox & MLB VR\\

  Moss Book II & VRChat & MVP Football - The Patrick Mahomes Experience\\

  Puzzling Places && Myth A Frozen Tale\\

  Red Matter 2 && Penn \& Teller VR F U, U, U, \& U\\

 Shave \& Stuff &&Prisms Math \\

  SOUL COVENANT && Space Explorers\\

  SWARM 2 && Spaceteam VR\\

  Swarm && Star Trek Bridge Crew\\

 The Room VR A Dark Matter && The Signifier\\

 Titans Clinic &&VR Karts Sprint\\ 

  Walkabout Mini Golf && VR Ping Pong Pro\\

  Hyper Dash && Zuma Legend VR\\

\bottomrule
\end{tabular}
\end{table}


\end{document}